# Effect of anomalously high oscillations of velocity of longitudinal ultrasound in high pure type I superconductor at weak external magnetic field

A. G. Shepelev, O. P. Ledenyov and G. D. Filimonov

*National Scientific Centre Kharkov Institute of Physics and Technology,
Academicheskaya 1, Kharkov 61108, Ukraine.*

We have experimentally observed, for the first time, the very strong oscillations of the velocity of the longitudinal ultrasound attenuation, $\Delta S/S = 0.6 \cdot 10^{-1}$, in a normal metal state of the high pure *type I* superconductor at the magnetic field $H$ the low temperature of *0.4 K*. The oscillations appear at $\omega \cdot \tau \sim 1$, following the change of either: the magnitude of the magnetic field $H$ or the orientation of the magnetic field $H$ in the plane $H \perp k$.

## Introduction

There have been many theoretical and experimental researches on the velocity of longitudinal ultrasound propagation (*US*) in the superconductors at the external magnetic fields. The biggest known oscillations of the ultrasound velocity are $\Delta S/S \sim 10^{-4}-10^{-3}$ (see the reported experimental investigations on the magneto-acoustic ("*Pippard*") oscillations [1], quantum oscillations [2] and giant quantum oscillations [3] at the magnitudes of the external magnetic field $H$ of $\sim$ *1, 70, 150 kOe* respectively).

## Discussion on measurement results

We report the results of experiments in which we observed oscillations $\sim \Delta S/S = 0.6 \cdot 10^{-1}$ in the normal metal state of the high pure *Gallium* single crystal at the external magnetic field of $\approx 50\ Oe$.

*1*. The effect is so large, that the time variation of the position of the *US* impulse, propagating through the sample, is observed directly on the screen of the oscilloscope, used in the experimental measurements setup to research the *US* absorption [4]. This has made it possible to conduct the high precise measurements at the frequencies of *30-90 MHz* by introducing only small complications in the apparatus. The propagated ultrasonic signal from the receiver's output was fed simultaneously to the pulse selector, the oscilloscope, and a stroboscopic attachment, *S1-21*, operating jointly with the oscilloscope, *S1-19B*. The use of the latter with the rigid synchronization of the entire measurement system allowed to measure the change of the position of the *US* pulse propagating through the sample relative to the reference signal. The measurement accuracy, using a sweep of *300 nsec*, is not worse than $\pm$ *6 nsec*, corresponding, at a sample's length *2 cm*, to the velocity change $\Delta S/S = \pm 1 \cdot 10^{-3}$. The use of the pulse selector and the *PDS-021 x-y* recorder makes it possible, when measuring the *US* velocity, to maintain the constant pulse amplitude with the accuracy $\pm 0.03\ dB$ at the receiver output. This has eliminated the errors, connected with the change of the *US* absorption in a sample at the external magnetic field $H$. The velocity measurements were conducted at the level of *1/3* from the main level of the pulse, and did not depend on the front of the pulse, used for the measurements.

The sample, synthesized at the *Experimental Plant of Rare Metals Institute* (*Giredmet Experimental Plant*), is the high pure *Ga* single crystal with the cylindrical shape with the diameter of *7 mm* and the length of *19.6 mm*. The cylinder's axis coincides with the crystallographic axis *b* in the high pure *Ga* single crystal. The *x*-cut quartz single crystals with the thickness of *0.3 mm* and the diameter of *4.5 mm* are attached to the end faces of a sample, and serve as the transmitter and receiver of the longitudinal ultrasonic signal. The *US* wave vector $k$ is oriented along the axis of the sample, which is in contact with the liquid *Helium-3* ($^3He$) in the cryostat [6]; the temperature in the cryostat can range from *0.4 K* to *2.2 K*. The external homogeneous magnetic field $H$ of a pair of *Helmholtz* coils, $H \perp k$, is oriented in the plane of the axes *a* and *c* of the high pure *Ga* single crystal with the accuracy $\pm\ 0.03°$. The vector $H$ can rotate in this plane with the speed of *1* revolution per hour. The magnetic field stabilization with the accuracy better than $10^{-5}$ and the variation of magnetic field's magnitude at the rate of *0.6 Oe/min* were effected by the means of the stabilization and sweeping systems. The *Earth's* magnetic field was cancelled out by the two pairs of *Helmholtz* coils.

*2*. At the fixed magnitude of the magnetic field $H > H_C \approx 50\ Oe$ and the temperature of *0.4 K*, we investigated the dependence of the propagation of longitudinal ultrasonic signal with the frequency of *70 MHz* on the orientation of the magnetic field $H$ in the plane of the axes *a* and *c* in the high pure *Ga* single crystal (the rotation diagram) [2]. It was observed that the strong oscillations of the ultrasound velocity, $\Delta S/S = 0.6 \cdot 10^{-2}$, appear in the certain region of the rotation diagram (Fig. 1), which accompany the oscillations of the ultrasound absorption $\Delta\Gamma$ (recorded with the *PDS-021 x-y* recorder). The maxima of the oscillations of the ultrasound velocity $\Delta S/S$ correspond to the minima of the oscillations of the ultrasound absorption $\Delta\Gamma$. The oscillations of the ultrasound velocity have a clearly defined region of the angles $\angle H$ between *H* and *a*, namely 36° - 48° (there are four angle regions, which are symmetrical in relation to the axis *a*, when the magnetic field, $H \gtrsim 50\ Oe$, is rotated on the angle of *360°*) and appear in the magnetic fields $H \sim 100\ Oe$. No such changes of oscillations of ultrasound velocity $\Delta S/S$ were observed outside these regions in spite of the fact that the ultrasound absorption $\Gamma$ changed strongly. The observed oscillations of ultrasound velocity $\Delta S/S$ appear at the low temperatures of *0.4 - 2.2 K*. The deviation of the rotation plane *H* from the condition $H \perp k$ by the angle of *1-2°* likewise does not result in a disappearance of the effect.

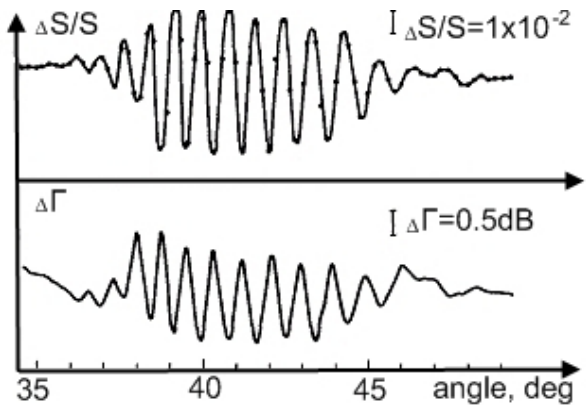

*Fig. 1. Dependence of oscillations of ultrasound velocity $\Delta S/S$ and oscillations of ultrasound absorption $\Delta\Gamma$ at frequency of 70 MHz on orientation of magnetic field H = 48.6 Oe in plane of axes a and c in high pure Gallium single crystal at **k** ∥ **b**, **k** ⊥ **H**, T = 0.4°K.*

*3.* At the fixed orientation of the magnetic field ***H***, in the indicated range of the angles $\angle H$ between ***H*** and ***a***, the variation of its intensity also causes the oscillations of ultrasound velocity $\Delta S/S$ of the same scale in Fig. 2, accompanied by the oscillations of ultrasound absorption $\Delta\Gamma$. The period of the oscillations is $\Delta(1/H) \approx 5 \cdot 10^{-4}\ Oe^{-1}$ at the frequency of *70 MHz*[3]. It should be noted that with the further increase of the magnitude of the magnetic field *H* there are the oscillations of ultrasound absorption with the similar amplitude (but a different period) $\Delta\Gamma$, which are not, however, accompanied by the big oscillations of ultrasound velocity $\Delta S/S$.

*4.* The picture of the phenomenon strongly depends on the frequency (wavelength) of the *US*. As it is seen in Fig. 3, the decrease of the frequency of the *US* by the factor of *2.3* results in the same increase of the period of oscillations; at the same time, the scale of the oscillations of ultrasound velocity $\Delta S/S$ decreases by the factor of *3*. Thus, these are not the quantum oscillations.

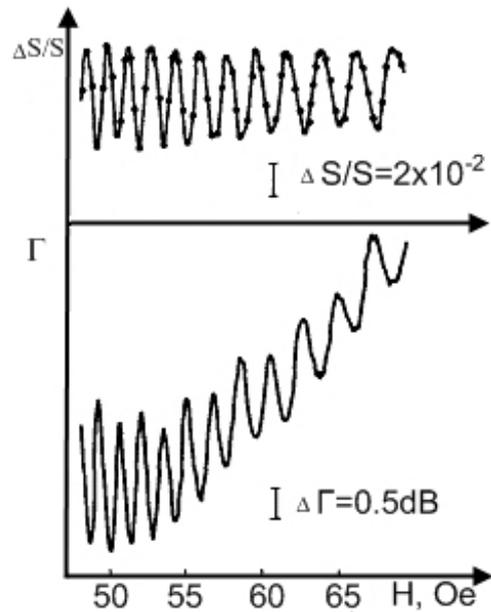

*Fig. 2. Dependence of oscillations of ultrasound velocity $\Delta S/S$ and oscillations of ultrasound absorption $\Delta\Gamma$ at frequency of 70 MHz on magnitude of magnetic field H at angle between H and a $\angle H,a = 40°$ in high pure Gallium single crystal at **k** ∥ **b**, **k** ⊥ **H**, T = 0.4°K.*

In light of the existing concepts, [9] at $\omega \cdot \tau \sim 1$ (where $\omega$ is the frequency of the *US* and $\tau$ is the electron relaxation time), in the strong magnetic fields $(k \cdot R \ll 1$, where *R* is the radius of the electron orbit in the magnetic field), there should be the monotonic change of the *US* velocity only[4]. In the weak magnetic fields $(k \cdot R \sim 1)$, the *US* velocity in the pure metal $(l \sim R$, where *l* is the electron mean free path) should experience the "*Pippard*" oscillations due to the change of the $k \cdot R$ at the influence of the magnetic field ***H***. Moreover, in so pure metal as the *Gallium*, the change of the frequency of the *US* can lead to the change of the velocity of the *US* not only because of the change of $\omega \cdot \tau$ in the "*Pippard*" oscillations, but also probably because of the change of the ratio $\omega/\Omega$, owing to the acoustic cyclotron resonance ($\Omega$ is the cyclotron resonance frequency).

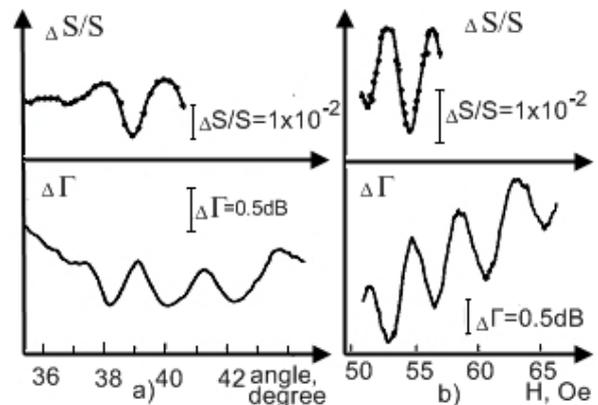

*Fig. 3. Dependences of oscillations of ultrasound velocity $\Delta S/S$ and oscillations of ultrasound absorption $\Delta\Gamma$ at frequency of 30 MHz at **k** ∥ **b**, **k** ⊥ **H**, T=0.375°K as functions of (a) orientation of magnetic field **H** at H=56.6 Oe, and of (b) magnitude of magnetic field **H** at angle $\angle H,a = 38.1°$.*

The scale of the phenomenon, generally speaking, suggests that the strong electron-phonon interaction or the collective electron excitations are present. The difficulty of the physical analysis is determined: by the complexity of the *Fermi* surface of the *Gallium*, by the presence of many strongly differing dimensions of the electron orbits, and by the anisotropy of both the effective electron-phonon interaction and of the electron relaxation time. It is possible that the groups of extended electron orbits that pass near saddle (or conical) points of the *Fermi* surface play a special role. The orbits with the two saddle points were observed in the indicated region of the orientations of magnetic field *H* in the researches on the magneto-resistance of the *Gallium* [11]. One cannot exclude the possibility of existence of the magnetic breakdown in the *Gallium* in the weak magnetic field [12], although the discussion o this problem is beyond the scope of this article.

To extend the temperatures range, the researches on the observed phenomenon in the high pure *Ga* single crystal of the same orientation and purity with the thickness of *6 mm*, but of the different geometric shape, were conducted at the another cryostat with the different magnetic system. The measurements were performed at the frequency of *90 MHz*, as before, using the first transmitted pulse; and at the frequency of *30 MHz*, using the pulse that made the three propagations through the sample. It turned out that the phenomenon takes place in the same interval of angles ∠*H* between *H* and *a*, and in the same magnetic field interval (i. e., it is not connected with the size effects) at the temperatures of *1.5-2.7 K*. At the higher temperature, this phenomenon is not observed, the fact that can be attributed to the decrease of the electron mean free path *l*.

We note that, at the temperature of *1. 5 K* and up to the magnetic fields of *30 Oe*, where the *US* absorption changes very strongly in the indicated range of angles ∠*H* between *H* and *a*, the big changes of the rate of *ΔS/S* are not observed.

## Conclusion

For the first time, we observed the very big oscillations of the velocity of the longitudinal ultrasound, $\Delta S/S = 0.6 \cdot 10^{-1}$, in a normal metal state of the high pure *type I* superconductor at the low temperature. The oscillations appear at $\omega \cdot \tau \sim 1$, following the variation of either: the magnitude of the magnetic field *H* or the orientation of the magnetic field *H* in the plane *H* ⊥ *k*.

The authors thank Boris G. Lazarev, E. A. Kaner, M. I. Kaganov, V. L. Pokrovskil, V. F. Gantmakher, A. A. Slutskin, A. M. Grishin, and V. S. Edel'man for the interesting thoughtful discussions on the experimental results.

1) The oscillatory ultrasound absorption was observed in an intermediate state of the *Gallium* of this kind earlier in [5]; this phenomenon can appear in a *very* pure superconductor only.
2) In our communication [7], the rotation diagrams in Figs. 2 and 3 show the plots of the amplitude of the *US*, propagating through the sample, *A(Φ)*, rather than *Γ(Φ)*.
3) The *French* physicists [8] observed the oscillations of the absorption, *ΔΓ*, in the *Gallium* under similar experimental conditions. In their opinion, these are the "*Pippard*" oscillations due to the six-hole *Fermi* surface.
4) In [10], it was established that, in the strong magnetic field of *15 kOe*, the velocity of the longitudinal *US* in the high pure *Ga* single crystal is increased by $\approx 3 \cdot 10^{-2}$.

———————